\documentclass[prb,twocolumn]{revtex4}
\usepackage{psfrag}
\usepackage{amssymb,amsmath,amsthm}
\usepackage[dvips]{graphicx}
\usepackage{verbatim} %
\usepackage{color}

\definecolor{magenta}{rgb}{0.7,0,0.7}
\definecolor{orange}{rgb}{1,0.5,0}

\newcommand{\beq}{\begin{equation}}
\newcommand{\eeq}{\end{equation}}
\newcommand{\bea}{\begin{eqnarray}}
\newcommand{\eea}{\end{eqnarray}}
\begin{document}
\begin{abstract} A molecular dynamics study of a two dimensional
system of particles interacting through a Lennard-Jones pairwise
potential is performed at fixed temperature and vanishing external
pressure. As the temperature is increased, a solid-to-liquid transition occurs. When the melting
temperature $T_c$ is approached from below,  there is  a
proliferation of dislocation pairs and the elastic constant
approaches the value predicted by the KTHNY theory.  In addition, as $T_c$ is approached from above, the
relaxation time increases, consistent with an approach to
criticality. However, simulations fail to produce a stable
  hexatic phase using systems with up to 90,000 particles. A
  significant jump in enthalpy at $T_c$ is observed, consistent with either a first order or a continuous
  transition. The role of external pressure is discussed.
\end{abstract}
\title{Characterization of the Melting
    Transition in Two Dimensions \\ at Vanishing External Pressure Using
    Molecular Dynamics Simulations}
\author{Daniel Asenjo$^{1,4}$, Fernando Lund$^1$, Sim\'on Poblete$^2$, Rodrigo Soto$^1$,
  and Marcos Sotomayor$^3$}
\affiliation{\mbox{$^1$Departamento de F\'\i sica and CIMAT, Facultad de Ciencias
F\'\i sicas y Matem\'aticas, Universidad de Chile, Santiago, Chile} \\
$^2$Max Planck Institute for Polymer Research, Ackermannweg 10, 55128 Mainz, Germany       \\
\mbox{$^3$ Howard Hughes Medical Institute and Neurobiology Department, Harvard Medical School, Boston, MA, USA} \\
\mbox{$^4$Department of Chemistry, University of Cambridge, Lensfield Road,
Cambridge CB2 1EW, United Kingdom.}
}
\maketitle
\section{Introduction}

Melting of an infinite solid in two dimensions has been
  described as a process driven by a proliferation of thermally
  excited dislocation pairs in the Kosterlitz-Thouless-Halperin-Nelson
  and Young (KTHNY) theory~\cite{kt,hn,y,nelson}. The theory,
  formulated at vanishing external pressure, predicts
the existence of a new, ``hexatic'', intermediate thermodynamic phase.
While solids are characterized by long range translational and orientational order,
liquids only present short range order. The predicted hexatic phase presents long range
orientational order but lacks long range translational order. The KTHNY theory predicts
a second order
  phase transition from the crystalline to the hexatic phase at
  which point there is a universal jump of a normalized elastic
  constant from a finite value to zero. This first transition is
  followed by a second transition from the hexatic phase to the liquid
  phase at a higher temperature. The KTHNY theory continues to
generate interest, specially because increased numerical capabilities
and new experimental techniques currently allow for new and more
accurate testing of theoretical predictions. Indeed, there have been
numerous attempts at the verification, both experimentally and
numerically, of the KTHNY theoretical predictions, with mixed
outcomes.

On the experimental side, studies with colloidal particles have
provided evidence of two stage melting with an intermediate hexatic
phase \cite{murraywinkle,kusneretal,zahnlenkem,linchen}, and of
elasticity behavior in agreement with the KTHNY predictions
\cite{grunbergetal}. The observed transition, however, appears to be
first order \cite{marcusrice,linchen}. Similar results have been
obtained with diblock copolymers \cite{angelescu}. Recently,
  melting in two steps with an intermediate hexatic phase has been
  observed in monolayers of polycristalline colloidal films, but not
  in thin or thick multilayer films~\cite{pengetal}. Also recently,
but in a different context, dislocations have been directly
observed in graphene \cite{granature}, prompting a renewed interest on
the role of defects in this two dimensional material
\cite{njp,prb1,prb2,prb3,prb4}.

On the numerical side, molecular dynamics and Monte-Carlo
  simulations~\cite{toxvaert,allen} of systems with a small number of
  particles ($N$) broadly detected a transition where the number of
  dislocations proliferates, but failed to provide clear evidence for
  the nature of the observed transition. First order melting
  has been reported in the literature, \cite{toxvaert,strandburg,bakker,lu}, while other calculations support a continuous transition
  \cite{li,fernandez}.

  The critical properties of the KTHNY transition
  are  a consequence of the renormalization effect that small scale
  fluctuations have on large scale fluctuations. For this mechanism to
  be operative, well separated length scales must exist, suggesting a
  minimum size for numerical simulations in two dimensions of
  $10^4$. Indeed, Chen et al.~ \cite{chenetal} performed molecular
  dynamics simulations of a Lennard-Jones system with a varying number
  $N$ of particles. They found a metastable hexatic phase for systems
  with $N\geq 36,864$, but not for $N\leq16,386$. In all cases,
  simulations were performed at a significant external pressure, a fact that alters the dislocation generation mechanism: the interaction between the components of a dislocation pair tends to close it down, while the external pressure, for some orientations, tends to open it up. The whole process becomes one of thermal activation, much like nucleation, and the likelihood of having isolated dislocations---and an hexatic phase---increases. A subsequent study in terms of inherent structure theory showed consistency with the KTHNY theory \cite{somer}. More
  recently, a molecular dynamics study \cite{shiba} carried out at constant volume,
  involving 36,000 particles interacting through a Lennard-Jones
  potential also reported the presence of an hexatic phase between the
  solid and liquid phases.  However, phase coexistence in $NVT$
  ensembles precludes unambiguous interpretation of these results.

On a different vein, in a three dimensional continuum elastic solid,
dislocation loops drive a mechanical instability at a finite
temperature \cite{l,al}, at which point the shear modulus vanishes as
a function of reduced temperature, following a power law with an
exponent whose value is a function that is independent of the
microscopic details of the elastic solid. Numerical calculations of
super-heated Lennard-Jones crystals near the melting
transition in three dimensions show the appearance of dislocations as
the temperature is raised \cite{gumbschprl, argprl}. However,
  it is unclear whether these dislocations play a central role during
  the phase transition or are just a by-product of another
  transition-driving mechanism.

The present work used molecular dynamics (MD) simulations in the
  $NpT$ ensemble to study the melting transition in two dimensions. MD has been chosen in this occasion over the possible Monte Carlo alternative in order to study the time evolution of the system, especially its relaxation behavior near criticality. We
  found a single-step solid-to-liquid transition (as determined by the
  enthalpy change) when a vanishing external pressure was applied to
  the system, in contrast to a multi-step transition at high pressure
  like the one presented in \cite{chenetal} using the same number of
  particles. However, within a narrow temperature interval defining
  the solid-to-liquid transition, the monitored relaxation times,
  elastic constants, and the evolution in number of dislocations were
  all consistent with the KTHNY theory. We suggest that, because of the necessary
  interplay between many length scales, a stable
  hexatic phase will be unambiguously observed only in systems with
  at least $\sim 10^6$ particles at zero external pressure.

\section{Molecular Dynamics Simulations}

Molecular dynamics (MD) simulations were carried out using a
  parallel MD code developed ``in house" based on the libraries
  presented in~\cite{allen2}. Simulation systems comprised $N$
  identical particles of mass $m$ in two dimensions interacting through a pairwise
  truncated and shifted Lennard-Jones (LJ) potential,
  \beq
 \label{eq:LJ}
  \phi_{\text{LJ}}(r)=4\epsilon\left[\left(\frac{\sigma}{r}\right)^{12}-
    \left(\frac{\sigma}{r}\right)^{6}\right] -C
    \eeq
for inter-particle distances $r$ smaller than a cutoff radius $r_c$
    and 0 for $r\geq r_c$. The value of $C$ was chosen to ensure
    continuity of the potential. All simulations were performed using
    periodic boundary conditions with a fully-flexible cell in the
    $NpT$ ensemble defined by the non-Hamiltonian equations of motion
    described by Martyna, Tobias, and Klein~\cite{martynaetal}. The
    equations of motion were integrated using the 5-value Gear
    predictor-corrector algorithm.

All simulation parameters and monitored quantities are
  expressed in reduced units: $x=x^*\sigma$; $t=t^*\sqrt{m\sigma^2/\epsilon}$; $
  T=T^*\epsilon /k_B$, where $k_B$ is the Boltzmann constant; $E=E^*\epsilon$; and
  $p=p^*\epsilon/\sigma^3$ (in the case of Argon, $\epsilon
  = 0.0104\,e$V and $\sigma = 3.4$~{\AA}, leading to a a unit of temperature $T$ being
  120.6~K and a unit of pressure $p$ being 42~MPa). The integration
  time step was set to $0.0005\sqrt{m\sigma^2/\epsilon}$ (34~fs for
  Argon)~\cite{chenetal}, ensuring extended-energy conservation to
  0.005\% per million iterations. We also verified momentum
  conservation and monitored pressure and temperature throughout
  simulations. Thermostat and barostat frequencies $\omega_p$ and
  $\omega_b$~\cite{martynaetal} were set to values between 80 and 100,
  and 0.1 and 0.4, respectively. The cutoff radius was set to 4 and a
  Verlet-neighbours list was used with a radius of 5.9.

Initial conditions were set by initial positions corresponding
  to a perfect triangular lattice, or a perfect triangular lattice
  plus a randomly oriented displacement of 0.05 LJ units, as well as initial
  velocities given by a Gaussian distribution corresponding to each
  temperature (initial condition IC1). Otherwise, initial conditions were assigned by
  equilibrium values of positions and velocities obtained from a
  previous run with similar values of $T$ and $p$ (initial condition IC2).
  Most runs were carried out with $N=36,864$, with up
to $1.7 \times 10^7$ time iterations, and a smaller number of runs
with $N=90,000$.

To characterize the state of the system throughout simulations
  we monitored the time evolution of the system's enthalpy and
  computed the pair and orientational correlation functions $g(r)$ and
  $g_6(r)$, respectively~\cite{chenetal}. The Lam\'e parameters were also computed,
  as a function of temperature. The number of dislocation
pairs present in the crystal below the transition was examined
counting the number of nearest neighbors for each particle, and
through visualization with the aid of a Voronoi construction.

\begin{figure}[h!]
\includegraphics[width = \columnwidth]{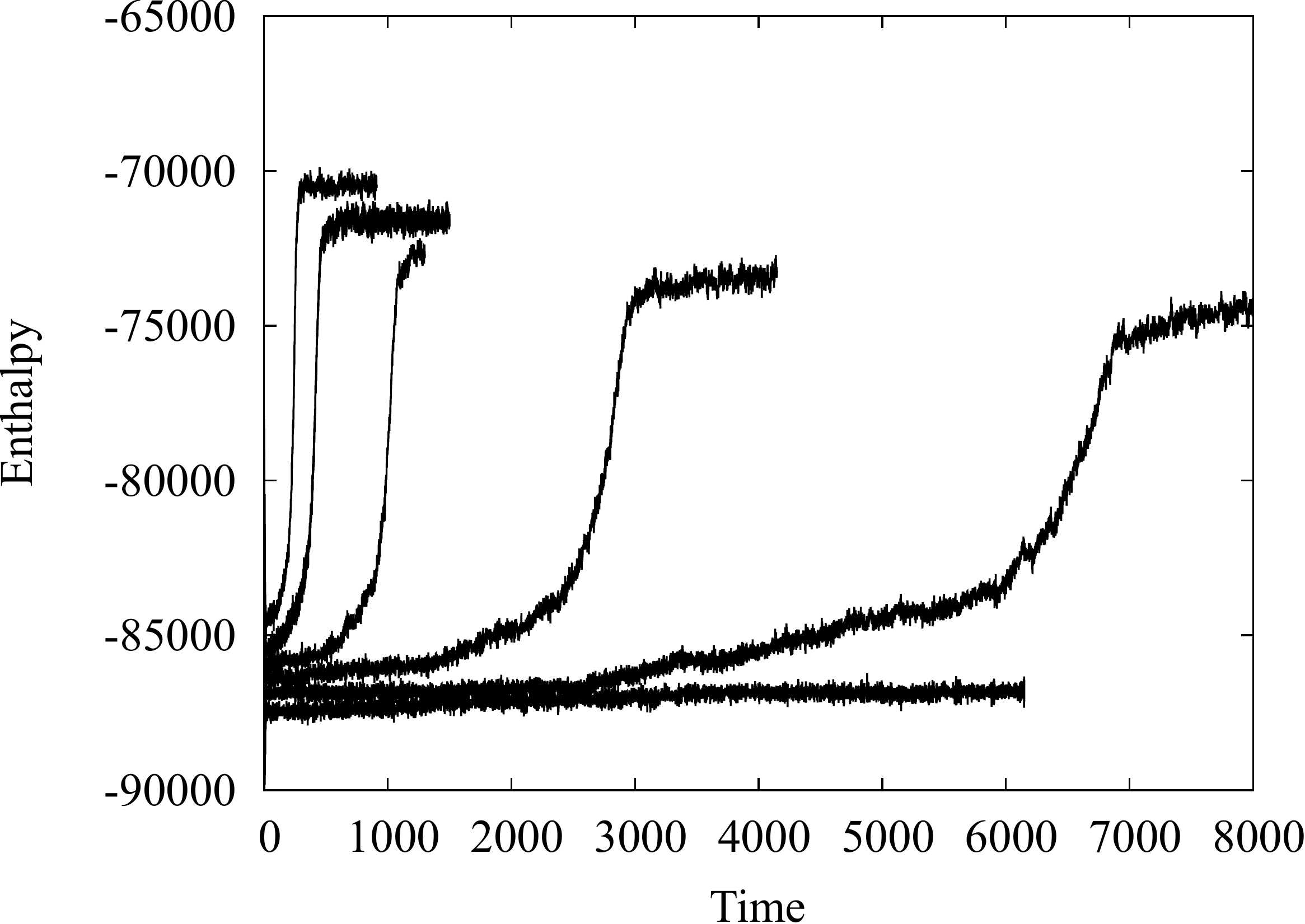}  
\includegraphics[width = \columnwidth]{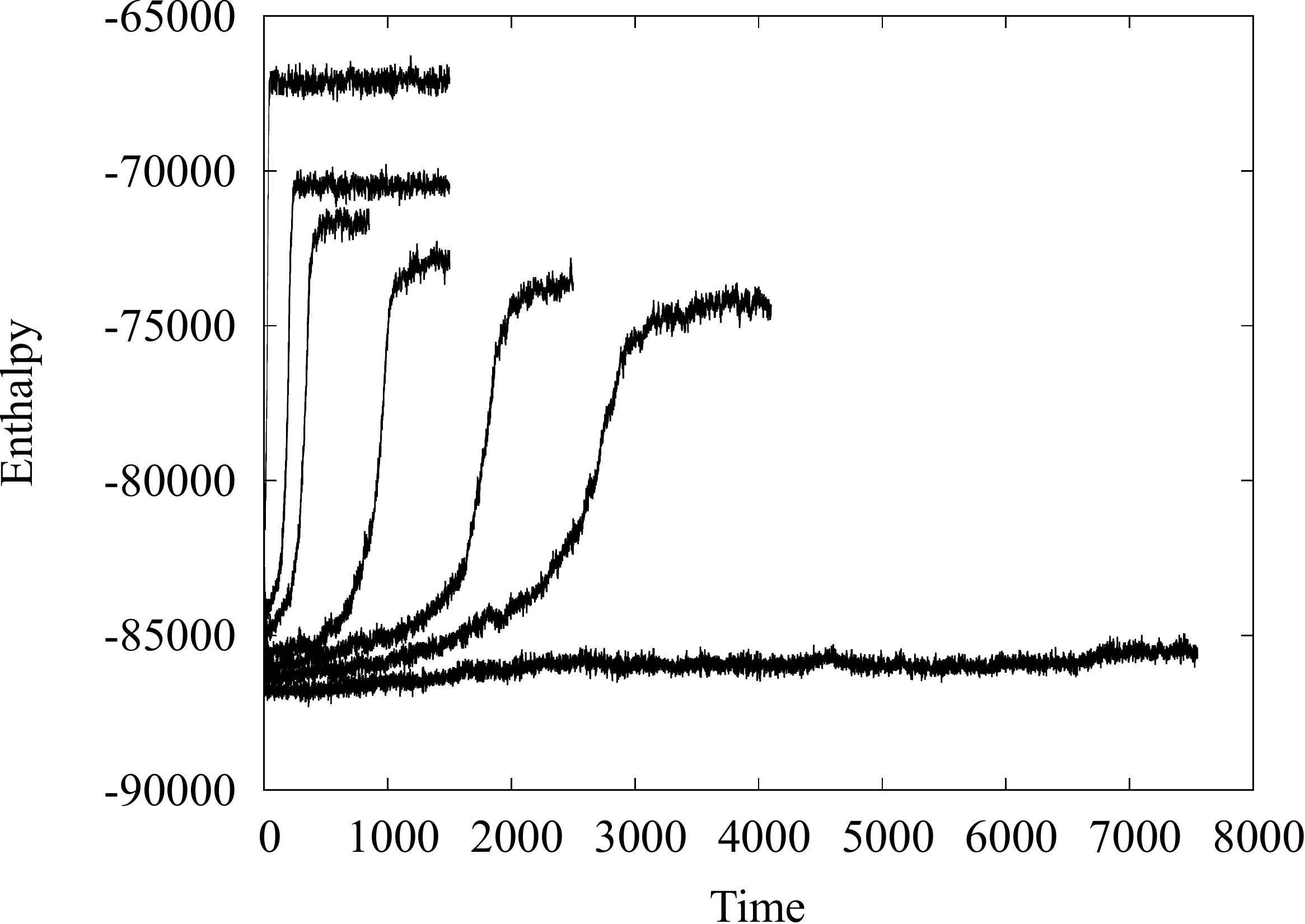}
\caption{Enthalpy as a function of time for several
  temperatures. Upper panel: initial conditions are given by initial
  positions corresponding to a perfect triangular lattice plus a
  randomly oriented displacement of 0,05 in LJ units, and initial
  velocities given by a Gaussian distribution corresponding to each
  temperature (initial condition IC1).  From the  bottom-right to the top-left curves the temperatures are T=0.4050, 0.4095, 0.4130, 0.4160, 0.4200, and 0.4250.Lower panel: positions and velocities are provided, at
  each temperature, by the equilibrium values obtained by a previous
  run at $T=0.40725$ (initial condition IC2). From the  bottom-right to the top-left curves the temperatures are T=0.40725, 0.41000, 0.41250, 0.41500, 0.42000, 0.42500, and 0.44000. All simulations were carried out at vanishing external pressure.}
\label{fig:hvst}
\end{figure}

\section{Results}

\subsection{Simulations at different temperatures show melting transition}

Two sets of simulations were performed to study the melting
  behavior of a LJ system. The first set of simulations, aimed at
  reproducing the results presented in~\cite{chenetal}, were
  performed at $p=20$ with $N=36,864$. Three simulations performed at
  temperatures $T_1=2.15$, $T_2=2.16$ and $T_3=2.17$ were carried out
  using a perfect crystalline lattice as initial condition. Despite
  the use of a different set of equations of motion that are modularly
  invariant, we obtained results that are consistent with those
  presented in~\cite{chenetal}. The enthalpy $h$ as a function of time
  remained stable during the simulation at $T_1$, increased rapidly in
  an apparent single step to achieve a stable value during the
  simulation performed at $T_3$, and increased in two steps with a
  transient state in the simulation performed at temperature $T_2$
  (data not shown). The pair and orientational correlations functions,
  $g(r)$ and $g_6(r)$, were consistent with a solid phase for the
  simulation performed at $T_1$ and with a liquid phase at the end of
  simulations performed at $T_2$ and $T_3$. The transient state
  observed at $T_2$ exhibits long-range orientational but not
  translational order, consistent with an hexatic phase. Similar
  results were obtained with a simulation performed at $T_2=2.16$ that
  used a thermalized (at $T=2.15$) set of initial conditions. These results
  show the possible existence of a metastable and transient hexatic
  phase when simulations are performed at a high external pressure
  $p=20$, as presented in~\cite{chenetal}.

\begin{figure}[h!]
\begin{flushright}
\includegraphics[width = .925\columnwidth]{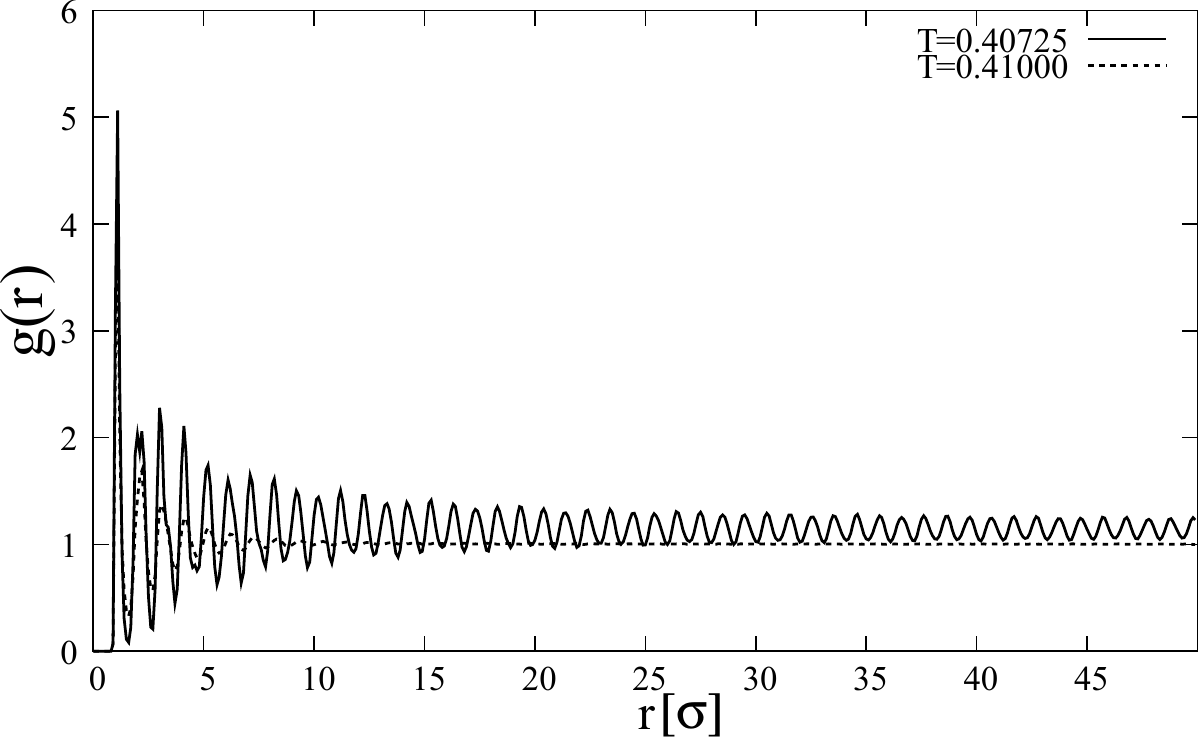}  
\includegraphics[width = \columnwidth]{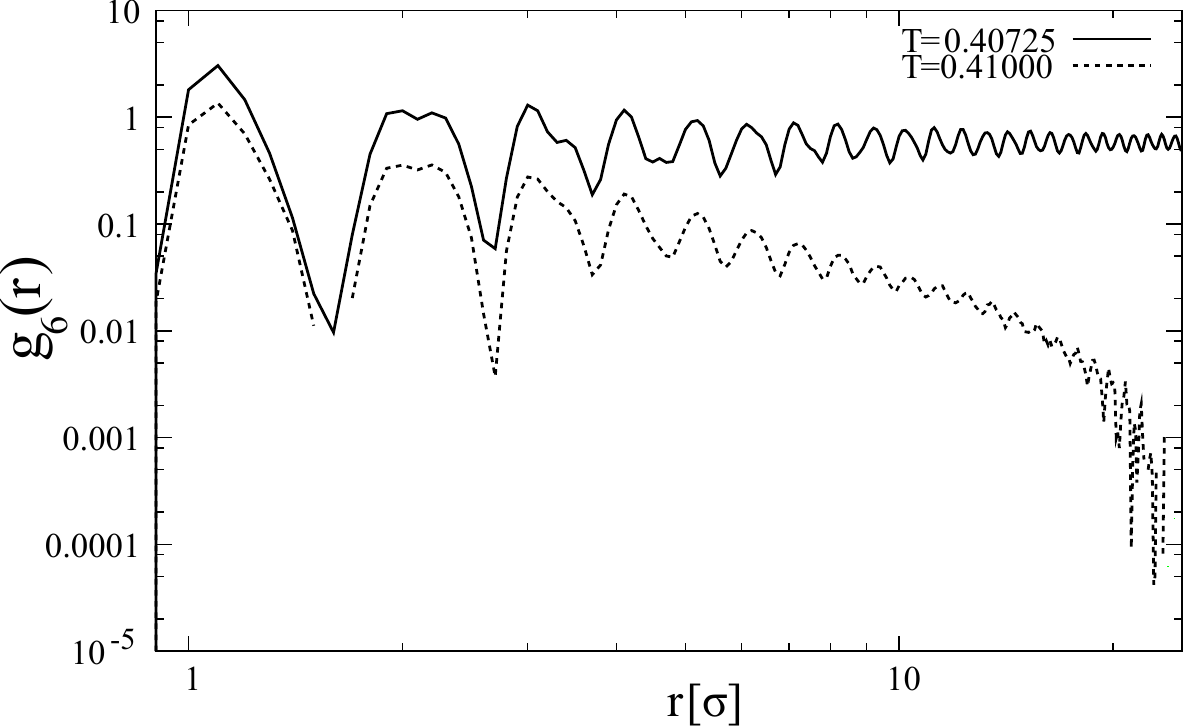}
\end{flushright}
\caption{Upper panel: Pair correlation function at
$T=0.40725$ indicates long-range translational order, and at $T=0.41000$, absence of it. Lower
panel, orientational correlation function at $T=0.40725$ indicates long range order, and
at $T=0.41000$, lack of it.}
\label{fig:hvsT2}
\end{figure}

\begin{figure}[h!]
\includegraphics[width = \columnwidth]{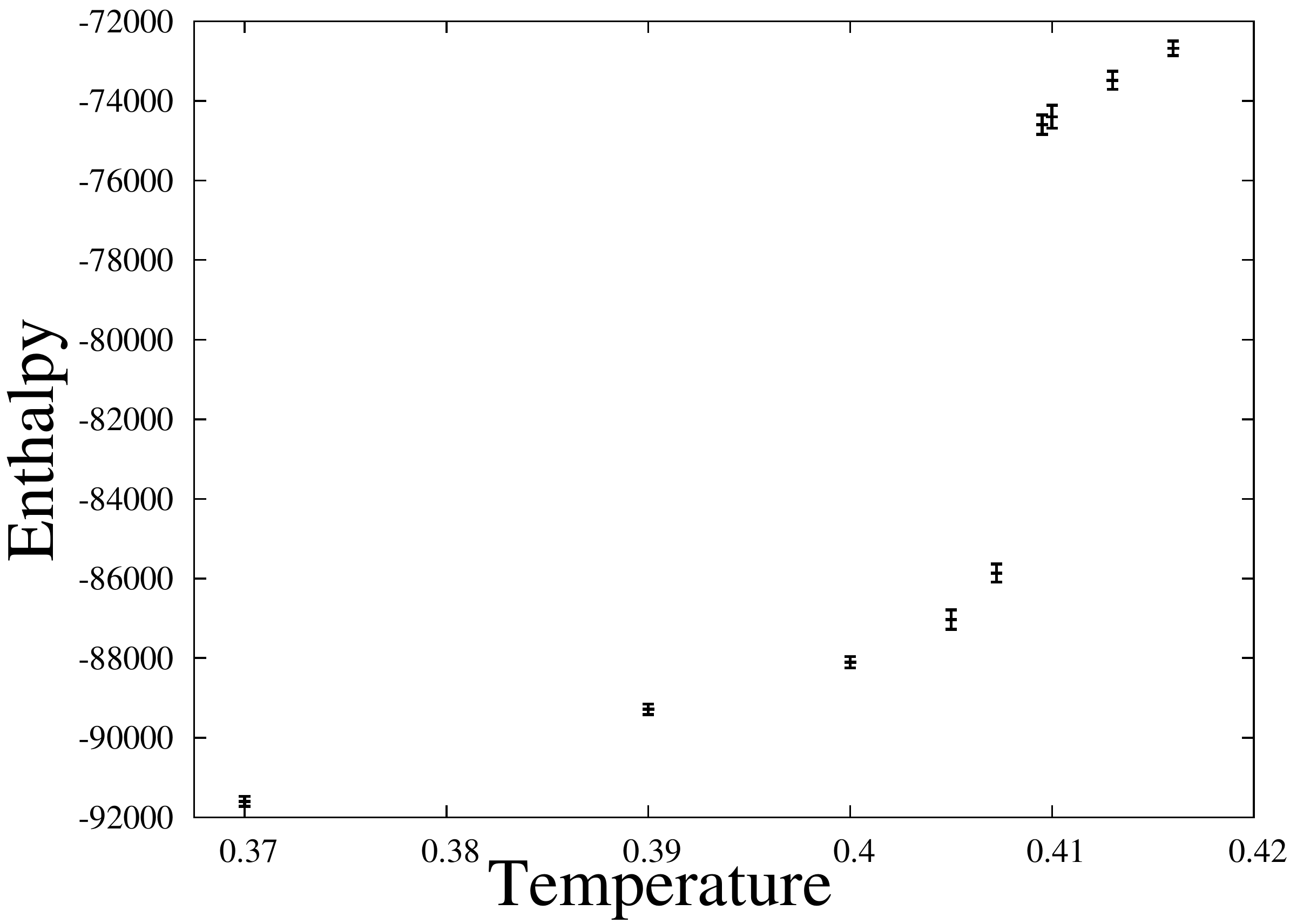}
\caption{Final enthalpy as a function of temperature. It presents a
  significant increase at $T_c = 0.40815 \pm 0.00090$. The low temperature phase is a
  solid, as evidenced by the behavior of pair and orientational
  correlation functions.  The high temperature phase is a liquid (Figure \ref{fig:hvsT2}).} \label{fig:hvsT1}
\end{figure}

To test the behavior of the system at a vanishing external
  pressure, we performed a second set of simulations in which a system
  with $N=36,864$ particles starting from initial conditions IC1 and IC2
   was simulated at several  different
  temperatures and pressure $p=0$ (more precisely, with vanishing
  normal and tangential stresses). As with the first set of
  simulations, the enthalpy $h$ of the system as a function of time
  (Figure 1) was stable at low temperature ($T\leq 0.40725$), but
  increased in an apparent single-step transition to an equilibrium
  value for high temperatures ($T\geq 0.4095$). The nature of the phase
  was characterized using $g(r)$ and $g_6(r)$ at the ending
  configurations of each simulation, confirming a liquid phase for
  $T\geq 0.4095$, and a solid phase for $T\leq 0.40725$ (Figure \ref{fig:hvsT2}). As opposed to
  the results obtained at $p=20$, no intermediate and transient state
  (possibly corresponding to a hexatic phase) was observed.  It is
  possible that a hexatic phase with an algebraically decaying
  $g_6(r)$ and an exponentially decaying $g(r)$ could be found within
  the interval of temperatures given by
  $T_-=0.40725<T<T_+=0.4095$. However, it was not possible to reach
  equilibrium in between these two temperature values within the
  available simulation time scales (data not shown). Defining the
  relative change in enthalpy as $\Delta h \equiv 2(h_{+}
  -h_{-})/(h_{+} +h_{-})$ (where $h_{\pm} \equiv h(T_{\pm})$) and the
  relative change in temperature as $\Delta T \equiv 2(T_{+}
  -T_{-})/(T_{+} +T_{-})$, we find $\Delta h = 0.1433$ and $\Delta T =
  0.0044$. Within this narrow temperature interval, our data is
  consistent both with an abrupt jump from a low to a high enthalpy
  value at some intermediate temperature, as would be the case for a
  first-order transition (i.e., with latent heat), as well as with a two-step change of enthalpy as a function of temperature within the
  temperature interval, as would be the case for a continuous
  transition, without latent heat (Figure \ref{fig:hvsT1}).

\subsection{Relaxation times slightly above melting increase as
the melting temperature is approached}
\begin{figure}[h!] 
\includegraphics[width = \columnwidth]{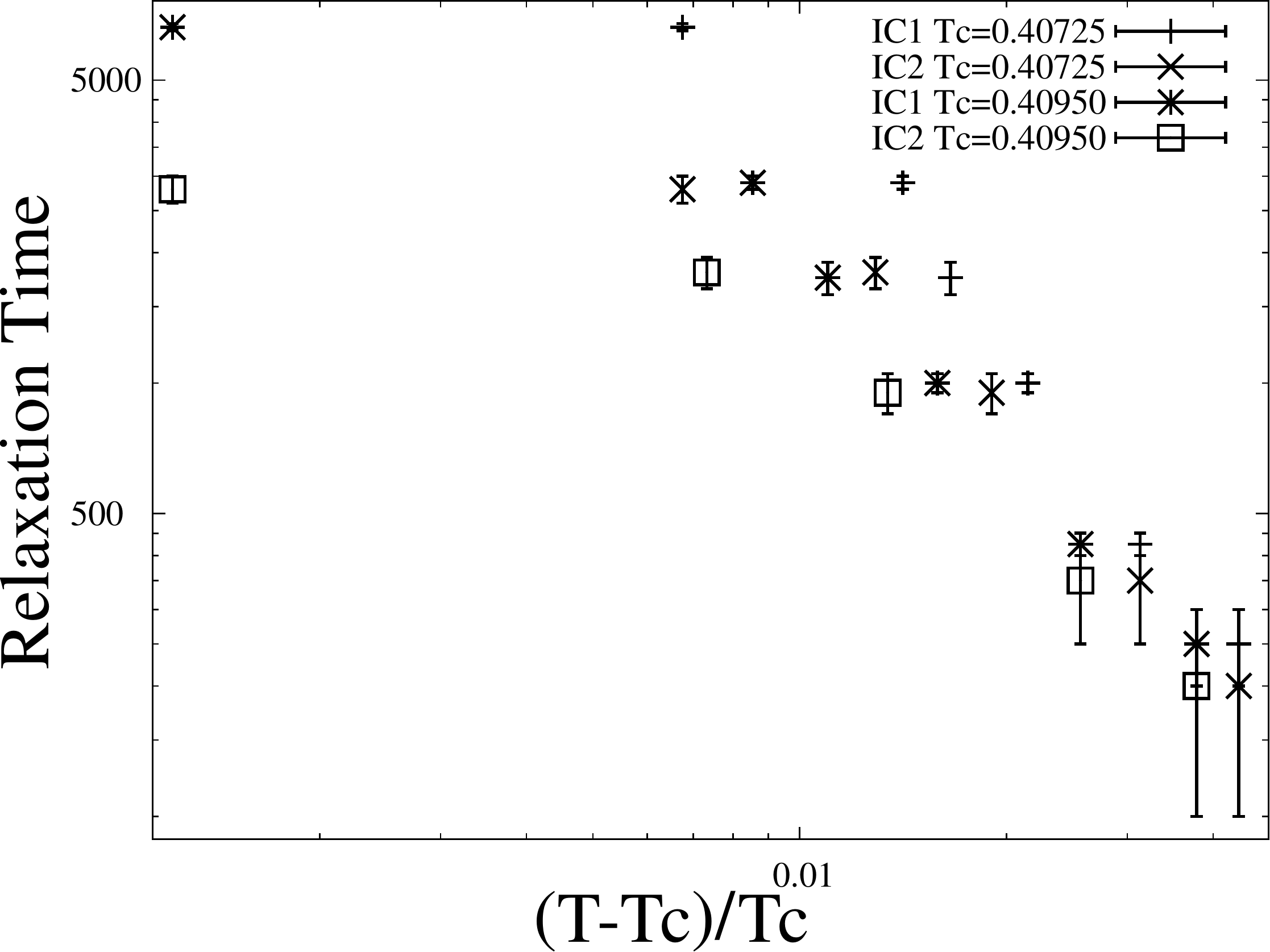}
\caption{Relaxation time as a function of temperature as the
  transition temperature $T_c$ is approached from above. There are
  four set of points, corresponding to the two different initial
  conditions (IC1 and IC2) indicated in Figure \ref{fig:hvst}, and two possible
  values for $T_c$, determined by the finite
  interval in which $T_c$ is found. The time increases without limit as $T_c$ is approached,
  consistent with criticality. Error bars are given by $t_{\rm min} - t_{\rm max}$.} \label{fig:time2eq}
\end{figure}

Near a critical point, relaxation times increase as criticality is
approached. This is due to the increasing size of fluctuations
\cite{hhrmp}, that reach macroscopic dimensions at the critical
point. To determine the nature of the phase transition
  observed at $p=0$, we monitored the relaxation time  ($t_R$) as a
  function of temperature in the simulations mentioned at $T>T_c$
  (Figure~\ref{fig:time2eq}).

  The relaxation time $t_R$ is defined, grossly, as the instant where the second derivative of enthalpy vs time vanishes, and more precisely as follows: the curve enthalpy vs time is interpolated by a smooth function whose second derivative is computed numerically, and the times $t_{\rm max}$, where curvature is a maximum and $t_{\rm min}$, where it is a minimum, determined. The relaxation time is then defined through $t_R = t_{\rm max} + (t_{\rm min} - t_{\rm max})/2$.

Within the accuracy of the simulation, this time grows without limit
as the melting temperature is approached, consistent with an approach
to criticality and a second order phase transition. The range of
values that was explored, however, was not large enough to detect a
possible power law behavior. These results did not depend on
  the initial conditions used (IC1 or IC2)
  and on the critical temperature used ($T_c=0.40725$ or
  $T_c=0.4095$, as there is no exact melting temperature but rather an
  interval $[T_-,T_+]$).

\subsection{Behavior of elastic constants is consistent with KTHNY theory}
Another way to determine the nature of the phase transition at
  $p=0$ and whether it is consistent with theoretical predictions
  consist on monitoring the behavior of the elastic constants of the
  system. The elastic response of an isotropic, homogeneous,
continuum solid is characterized by two Lam\'e coefficients $\lambda$ and
$\mu$. They appear in the compliance tensor as
\begin{equation}
  S_{ijkl}=\frac{1}{4\mu}\left(\delta_{ik}\delta_{jl}+\delta_{il}\delta_{jk}-\frac{\lambda}{\lambda+\mu}\delta_{ij}\delta_{kl}\right)
\end{equation}
where
\[
\epsilon_{ij} = S_{ijkl} \sigma_{kl} ,
\]
$\epsilon_{ij}$ is the strain and $\sigma_{kl}$ the stress. There are
several possible ways of extracting $\lambda$ and $\mu$ from this tensor.
In the continuum theory they are of course
equivalent. But in a numerical calculation involving a finite number
of atoms, this will no longer necessarily be the case. They should
coincide, however, within numerical accuracy and error bars (see below). The following relations hold
for the combination of Lam\'e coefficients:
\begin{eqnarray}
  K&\equiv&\frac{4\mu(\lambda+\mu)}{2\mu+\lambda} \\
  &=&\frac{1}{S_{0000}} \equiv  K_1 \\
  &=&\frac{1}{S_{1111}} \equiv  K_2 \\
  &=&\frac{1}{S_{0011}+2S_{0101}} \equiv  K_3
\end{eqnarray}

It is a significant prediction of KTHNY theory that $K$ approaches a
universal value as the critical temperature $T_c$ is approached from below:
\begin{equation}
  \lim_{T\to T_c^-}K=16\pi \frac{k_BT_c}{b^2} \equiv K_c
\end{equation}
where $b$ is the Burgers vector of the dislocations (given by the
lattice constant at zero temperature). $K$ vanishes above
$T_c$.

We have computed the strain of our system following Ray and Rahman \cite{ray,rayrahman},
\begin{equation}
  \mathbf{\epsilon}=\frac 1 2 \left[({\mathbf
  h}_R^{-1})^t{\mathbf h}^t{\mathbf
  h}{\mathbf h}_R^{-1}- {\mathbf I}\right]
\end{equation}
where ${\mathbf h}$ is a two-by-two matrix whose column vectors define
the simulation box, ${\mathbf h}_R$ is a reference box (here taken as
the time average of the simulation box) and ${\mathbf I}$ is the
identity matrix. The compliance tensor is given in terms of strain
fluctuations through
\begin{equation}
  S_{ijkl}=\beta V_R (\langle
  \epsilon_{ij}\epsilon_{kl}\rangle-\langle
  \epsilon_{ij}\rangle\langle \epsilon_{kl}\rangle)
\label{compliance}
\end{equation}
where $V_R$ is the volume of the reference box.

\begin{figure}[h!]
\includegraphics[width = .3\columnwidth]{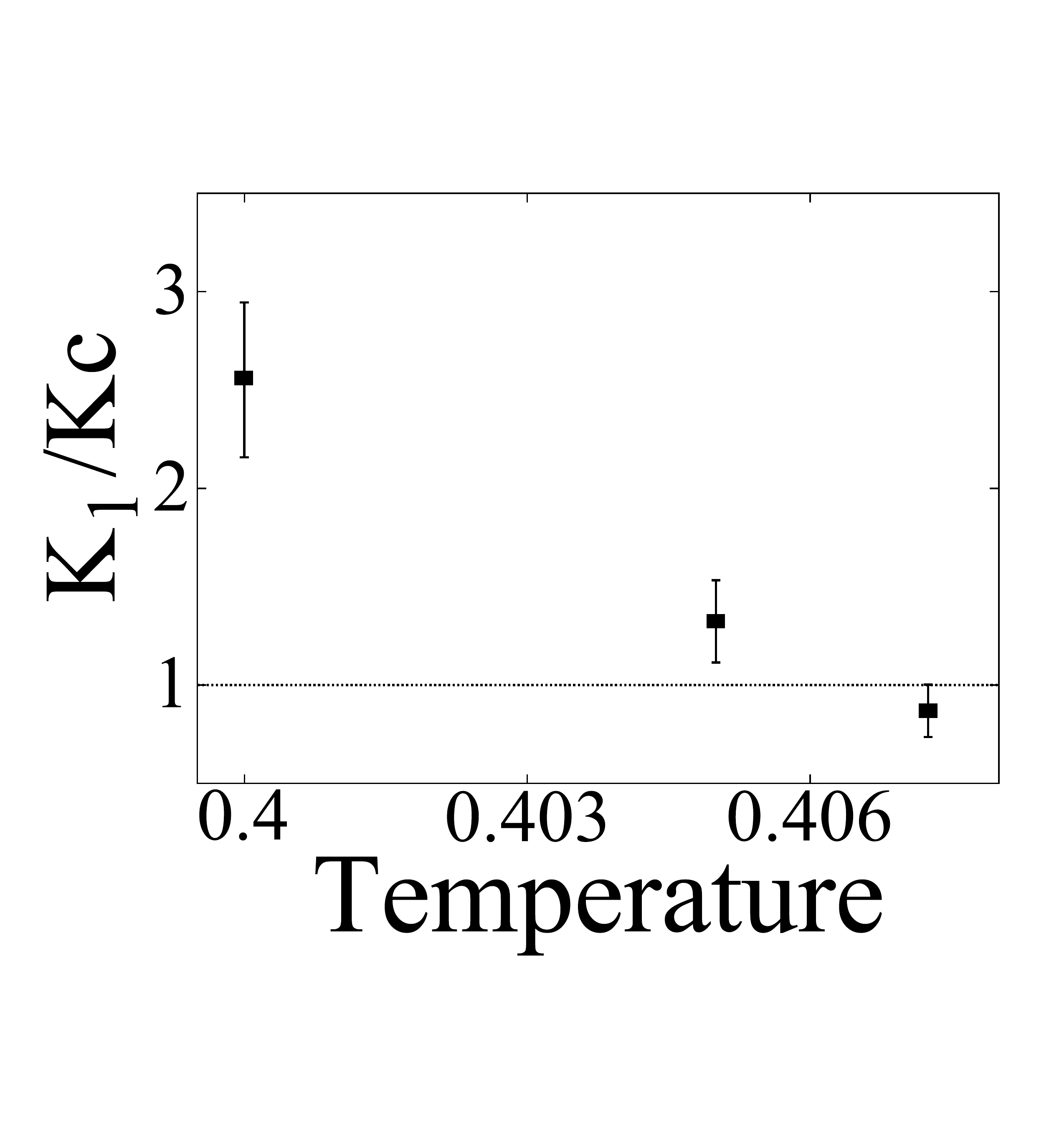} 
\includegraphics[width = .3\columnwidth]{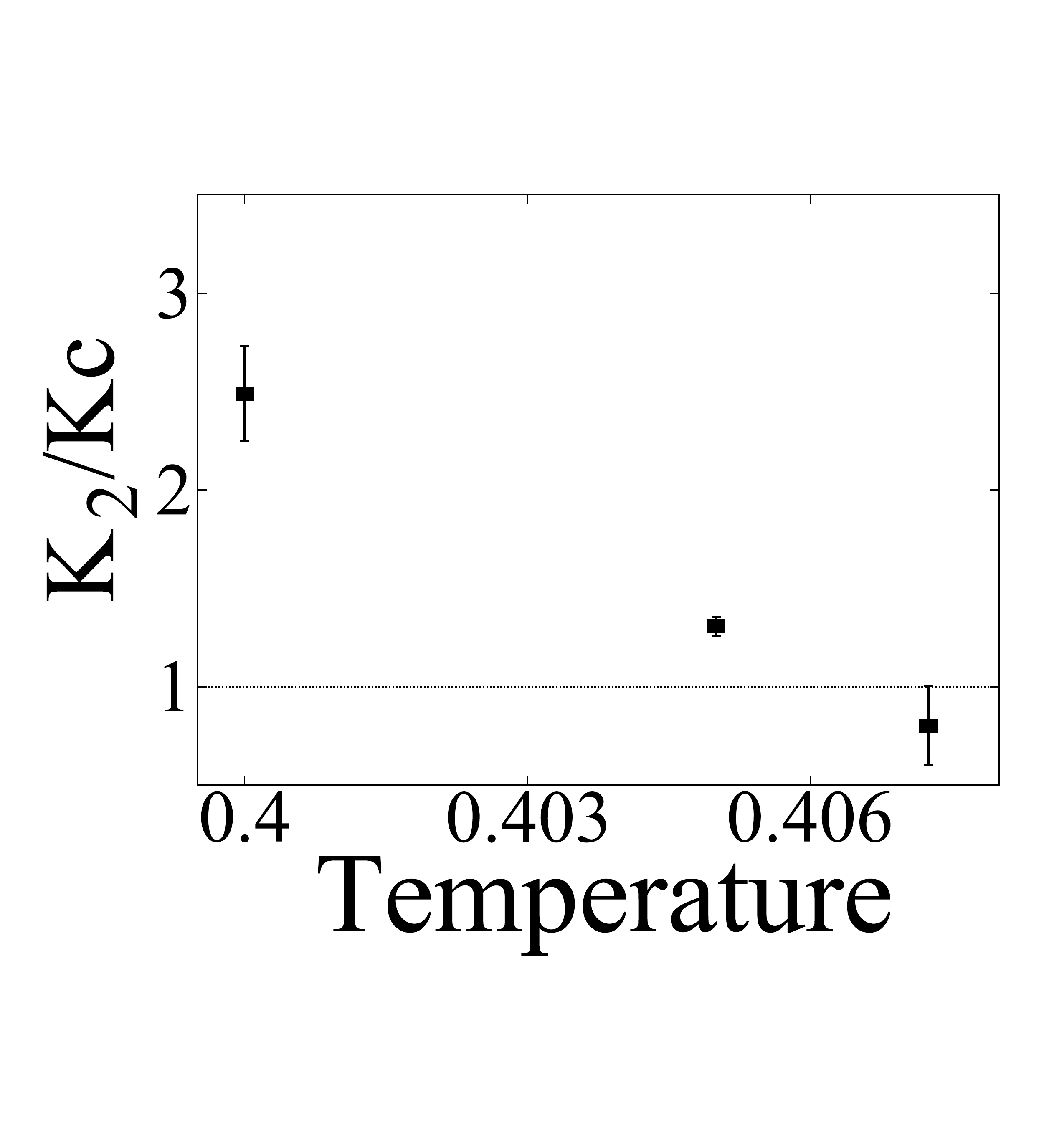} 
\includegraphics[width = .3\columnwidth]{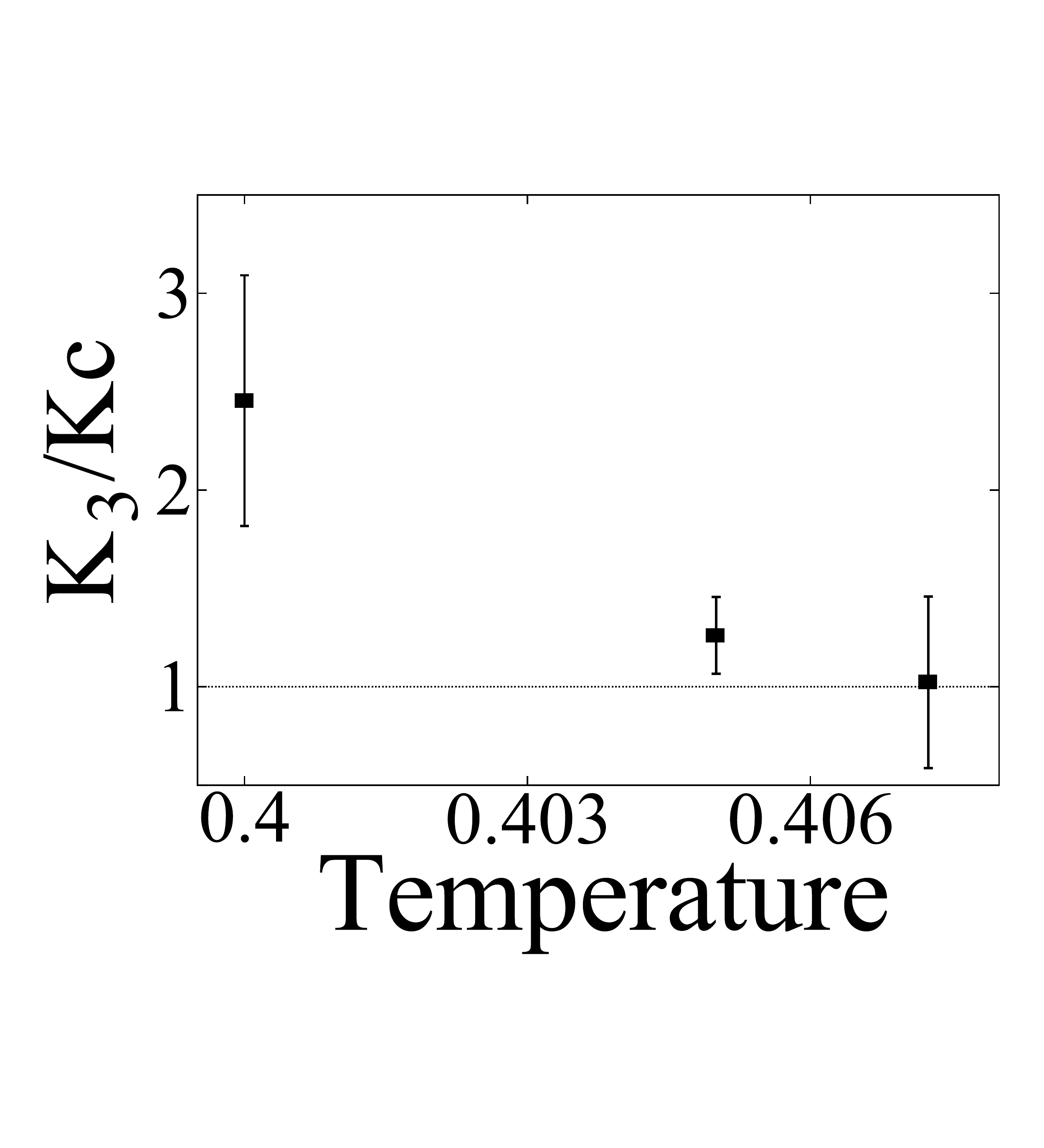}
\caption{Elastic constant $K_1/K_c$ (left-hand-side panel), $K_2/K_c$ (middle panel)
  and $K_3/K_c$ (right-hand -side panel) as a function of
  temperature. Within numerical accuracy they coincide, as they should. Near the transition their value is consistent with $1$, as predicted by KTHNY theory.}
\label{fig:elast_const}
\end{figure}

The compliances were calculated from simulations of $2.5\times 10^6$ time steps,
after $5\times 10^5$ equilibration steps. The error bars were estimated by performing
blocking averages \cite{frenkel, janke} and then propagating the error in
equation \ref{compliance}. The simulation data was divided into 5 data blocks.
The value of the Burgers vector was estimated as the lattice constant that minimizes
the energy of a triangular crystal, $b_0=1.11145 \sigma$.
Figure \ref{fig:elast_const} shows the ratios $K_1/K_c$, $K_2/K_c$ and $K_3/K_c$ as a
function of temperature as the transition is approached from
below. The computed values are consistent with the KTHNY theory.

\subsection{Proliferation of dislocations is consistent with KTHNY theory}
\begin{figure}[h!] 
 \includegraphics[width = \columnwidth]{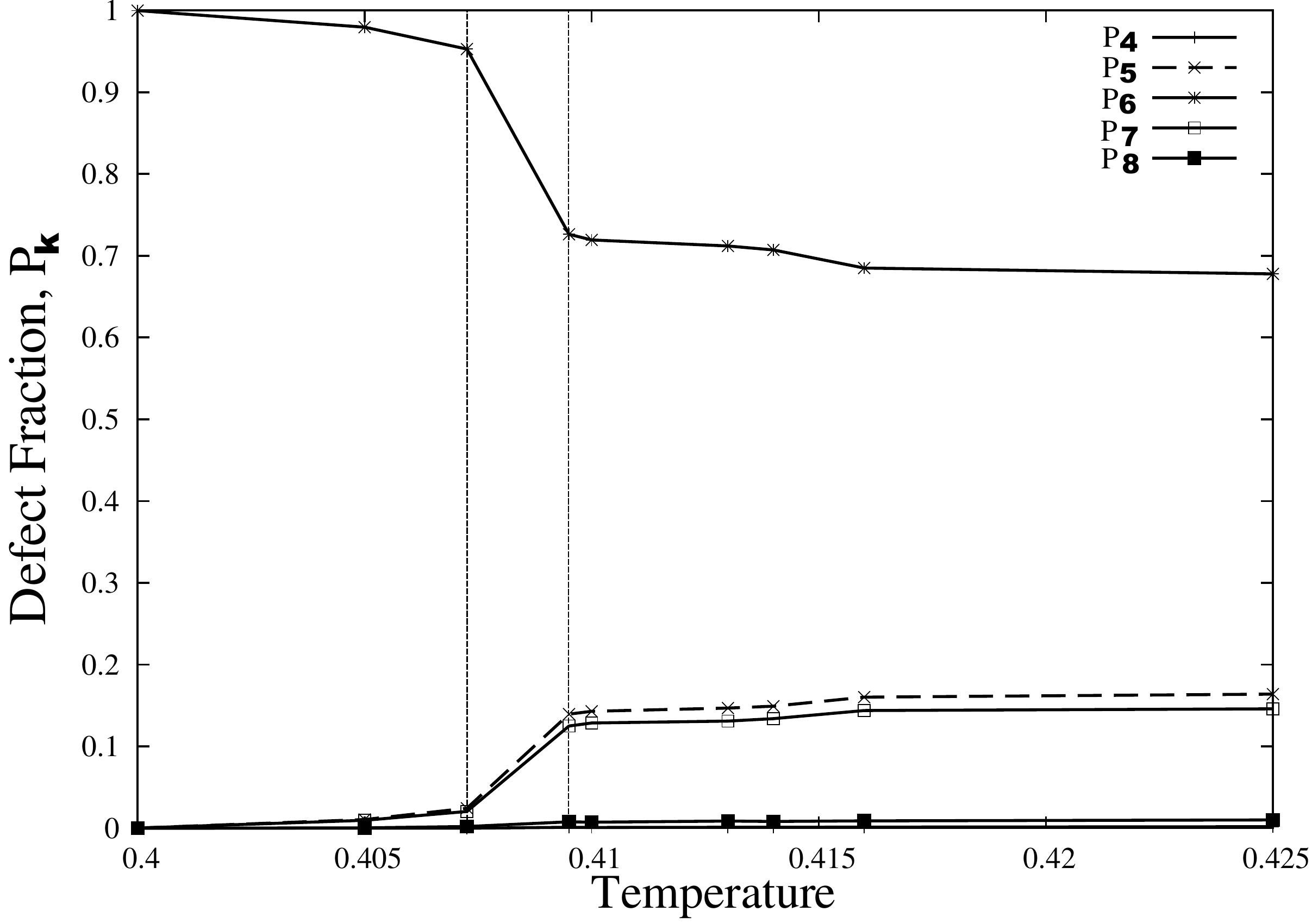}
\caption{$P_k$, the fractional number of sites with $k$ neighbors, as
  a function of temperature for $k=4,5,6,7,8.$ There is a significant
  decrease in $P_6$ across the solid-liquid transition with a
  corresponding increase in $P_5$ and $P_7$, consistent with a
  transition driven by the unbinding of dislocation pairs.}
\label{fig:hist}
\end{figure}

Finally, and to fully characterize the transition observed at $p=0$,
we monitored the number of dislocations as a function of temperature,
counting the number of neighbors for each point using a Voronoi
construction after equilibrium had been reached. Figure \ref{fig:hist} shows $P_k$, the
fractional number of sites with $k$ neighbors, as a function of
temperature. Of course, for a perfect triangular lattice, $P_6 =1 $
and $P_k =0$ for $n \ne 6$. A dislocation is characterized by two
neighboring sites, one with five, and the other with seven,
neighbors. The KTNHY theory predicts that the loss of long range
translational order is due to the proliferation, and subsequent
unbinding, of thermally generated
dislocation pairs. Such pairs will be characterized then by clusters
of four sites, two of them with five, and two of them with seven,
neighbors. Figure \ref{fig:hist} shows that across the solid-to-liquid
transition there is a significant decrease in $P_6$, and a
corresponding increase in $P_5$ and $P_7$, consistent with the KTHNY
theory.

\begin{figure}[h!] 
\includegraphics[width = .4\columnwidth]{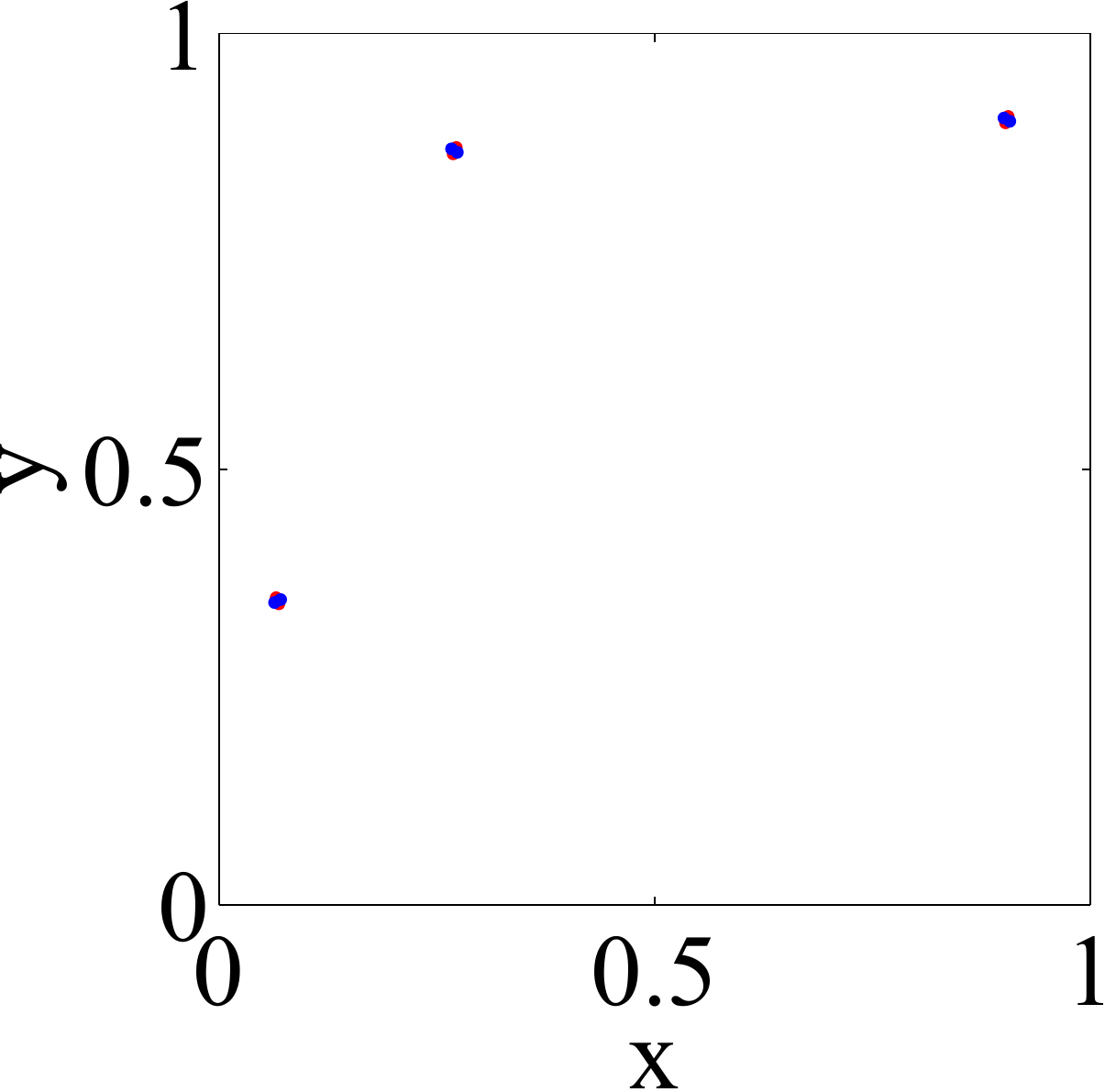}
\includegraphics[width = .4\columnwidth]{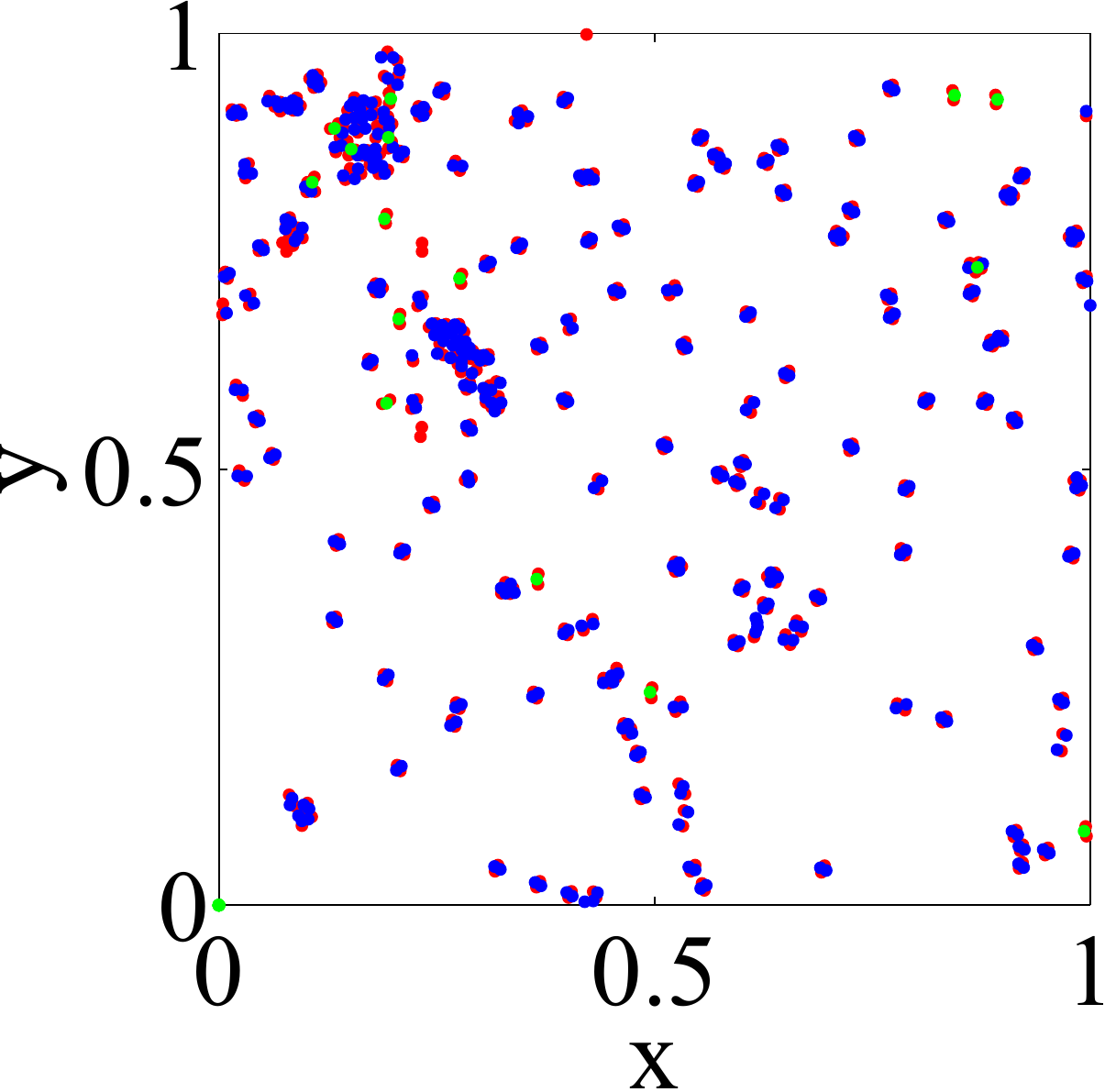}
\\ \includegraphics[width = .4\columnwidth]{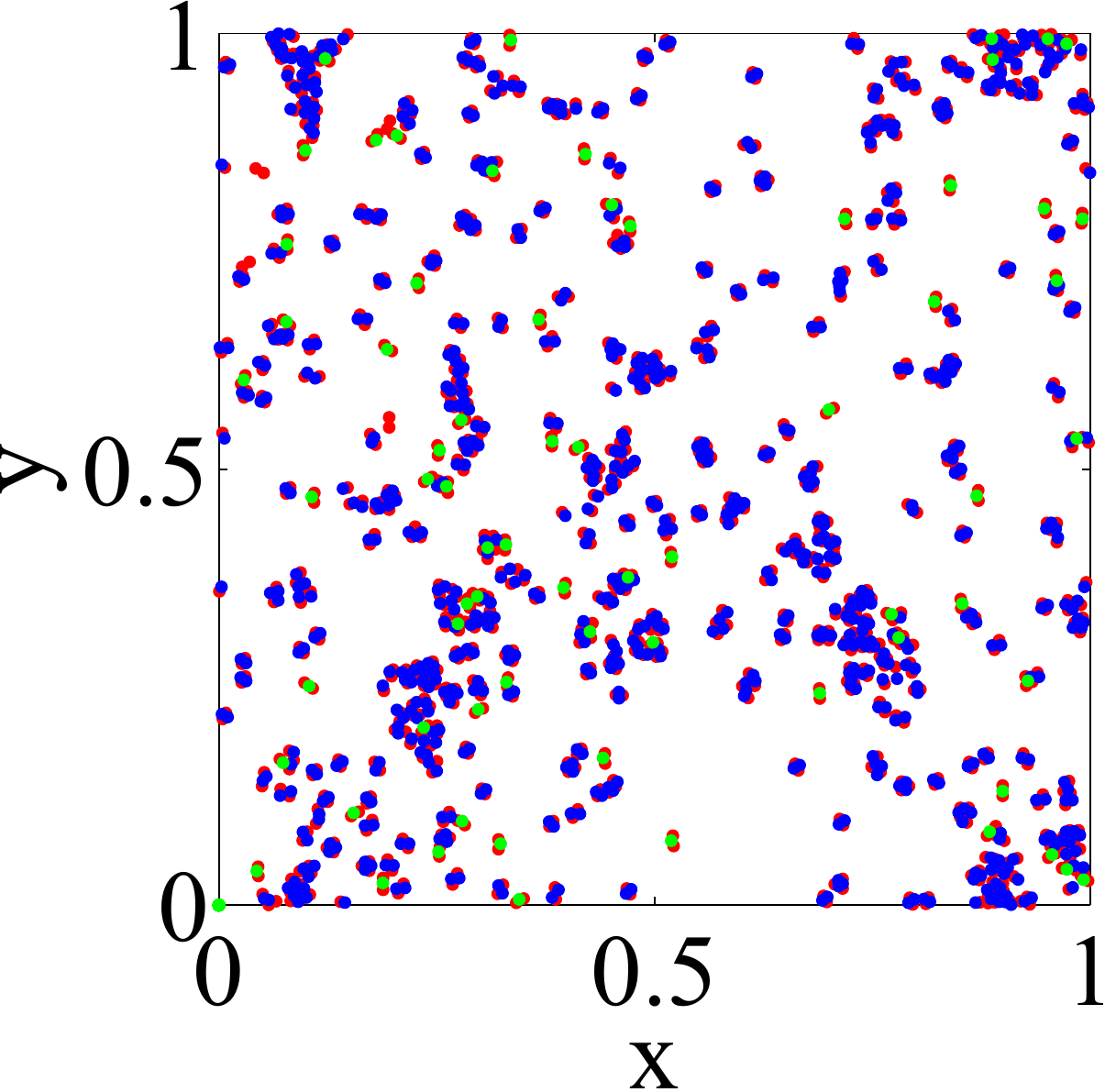}
\includegraphics[width = .4\columnwidth]{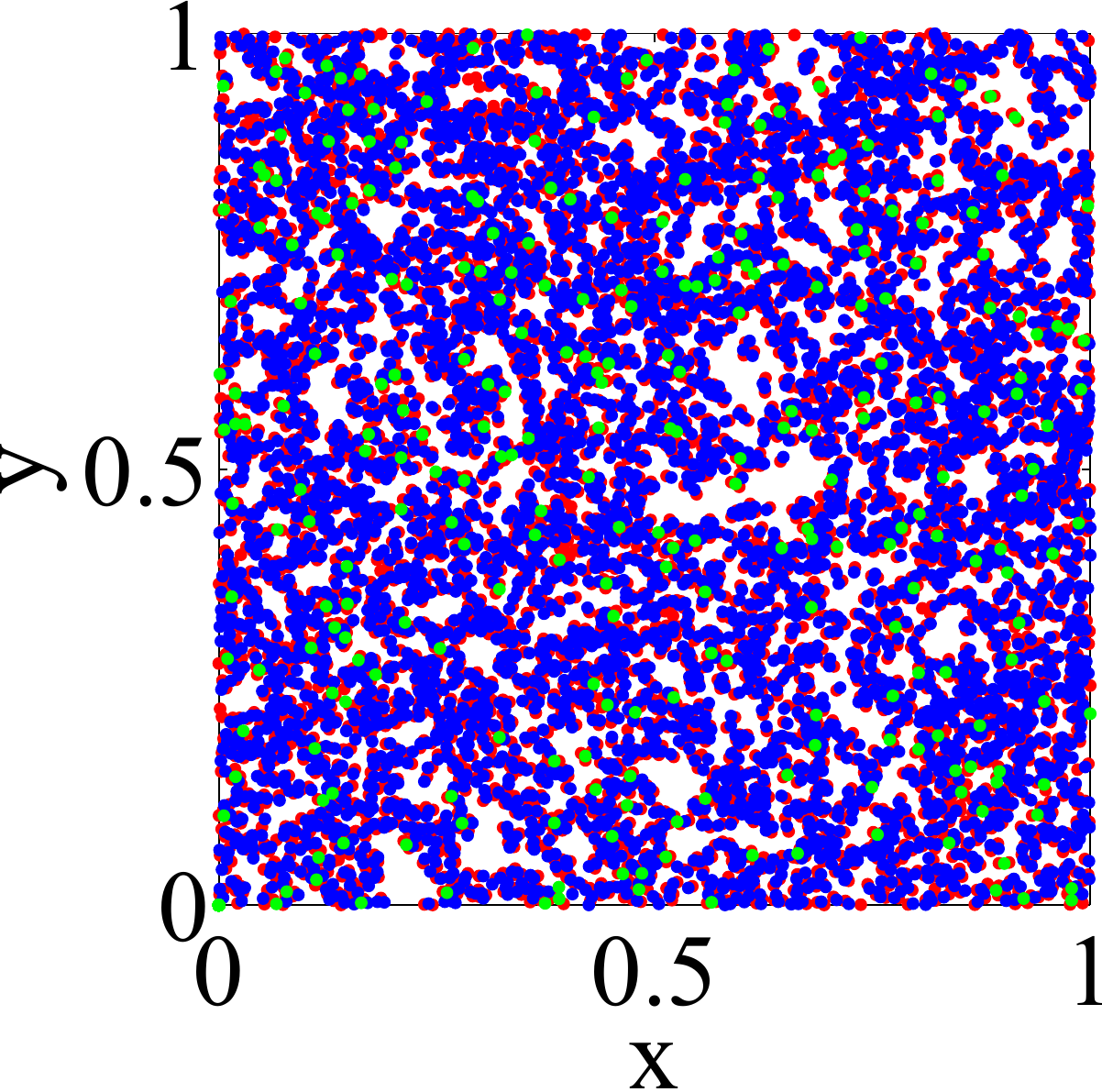} \\
\caption{Dislocation dipole population as a function of temperature, monitored
  with the sites having 5 (red) and 7 (blue) neighbors. Upper left panel, $T=0.400$;
  upper rigth  panel, $T=0.405$; lower left panel, $T=0.40725$, all three temperature values below $T_c$. Lower rigth panel, $T=0.4095$ just
  above $T_c$. The number of dislocation dipoles steadily increases as the
  transition is approached. Sites with 8 neighbors (green) are also indicated.
}
\label{fig:5-7}
\end{figure}

Figure \ref{fig:5-7} also provides a visual illustration of the number
of dislocations, monitored with the number of sites having 5 or 7
nearest neighbors, within the simulation box for different
temperatures. Their proliferation is apparent, consistent with the
KTNHY theory.

\section{Discussion: The role of external pressure}
An hexatic phase has been observed in simulations presented by
  Chen et al. \cite{chenetal} and also reproduced here. However, this
  phase is transient, not in equilibrium, and occurs when the
  Lennard-Jones system is subjected to a significant external pressure
  of 20. The use of a constant pressure ensemble molecular dynamics
ensure the existence of a homogeneous phase. However, we have been
unable to observe the hexatic phase at a constant vanishing external
pressure. Is there a reason, within the KTHNY theory, not to observe
an hexatic phase at zero external pressure with a finite number of
particles?

The hexatic phase arises \cite{hn} after (i.e., at a higher
temperature) a triangular lattice undergoes a dislocation unbinding
transition but before (i.e., at a lower temperature) a disclination
unbinding transition occurs. The latter is possible because the plasma
of free dislocations screens the disclination-disclination
interaction, allowing a transition much like the one originally
considered by Kosterlitz and Thouless \cite{kt}. As mentioned in the
Introduction, it is critical, in a numerical simulation, to have
several length scales available, since the KTHNY mechanism involves
the interaction among defects of different sizes, at many scales. This
is in addition to the fact that the theoretical analysis is carried
out in the thermodynamic limit. We now argue that at least 10$^6$ particles are needed in simulations, in order to have three decades in length scales.

The physics of dislocation unbinding changes considerably when an
external pressure $p$ is included. Indeed, the energy $U$ of a
dislocation dipole with Burgers vector $\vec b$, whose components are
separated by $\vec R$ in this case is \cite{hl} \beq U(R) = \frac{b^2
  K}{4\pi} \left[ \log \left( \frac{R}{\tau} \right) + {\cal C} -
  \frac 12 \cos 2 \theta \right] + P b R \sin \theta
\label{dislocdipol}
\eeq where $\tau$ is the dislocation core size; $\cal{C}$, roughly,
determines the core energy (i.e., the minimum energy needed to
generate a dipole); $R=|\vec R|$; $b=|\vec b|$ and $\theta$ is the
angle between $\vec R$ and $\vec b$.  Clearly, when $\sin \theta < 0$
the last term on the right hand side turns the dislocation-dipole
unbinding process into a thermally activated one, with an activation
energy $U_A$ (taking $\theta = -\pi/2 $, the most favorable case, for
illustration purposes)
\[
U_A \equiv \frac{b^2 K}{4\pi} \left[ \log \left( \frac 1{4\pi} \frac KP \frac b{\tau}
\right) + {\cal C} - \frac 12 \right] .
\]
Consequently, and given the logarithmic dependence on the ratio of
external pressure to elastic constant, even a modest value of external
pressure $P$, compared to $K$, will give values for the activation
energy in the same ball park as the chemical potential for the
dislocation dipole, and a proliferation of isolated dislocations will
ensue. Thus, it will not be surprising in those circumstances to
observe an hexatic phase. However, Eqn. (\ref{dislocdipol}) shows that, at any given non vanishing pressure there will be a finite rate of dislocation generation, driving the system away from equilibrium. A quantitative study of this interesting phenomenon is outside the scope of the present paper.

In the absence of external pressure, disclination pairs above the
dislocation unbinding transition interact via an energy that depends
on the logarithm of their mutual distance, with a coupling (called
$K_A$ by Nelson and Halperin \cite{hn}) that is finite due to the
screening effect of the free dislocations. A second transition towards
the liquid state thus occurs because of two distinct screenings: free
dislocations screen the interaction between disclination pairs to an
effective logarithmic interaction, and then this interaction is
renormalized because of the interaction between disclination pairs at
different length scales. So, three length scales should be needed for
this scale dependent interaction among disclination pairs to become
operative. In addition, a further length scale would seem to be
necessary in order to have enough free dislocations in between a
disclination pair for their interaction to be effectively
screened. According to this reasoning, at least $\sim 10^6$ particles
would be needed to observe an hexatic phase as an equilibrium phase at
zero external pressure in a numerical simulation.

\section{Conclusions}
The KTNHY theory of melting in two dimensions \cite{kt, hn, y}
involves the interaction among dislocation dipoles whose sizes span
different length scales. Thus, a numerical simulation that aims at
verifying the theory should involve enough different length scales for
this interaction to be possible. In two dimensions, $10^4$ particles
thus appears as an a absolute minimum.  Increasing this number should
improve the statistics. By the same token, $10^6$ particles should be
a minimum number to capture this type of effect in three
dimensions. Our simulation has been carried at constant (vanishing)
external pressure in order to prevent phase coexistence. We find a
solid-to-liquid transition, and the behavior of the solid phase as the
transition temperature is approached is consistent with the
predictions of KTNHY.  The behavior of enthalpy as a function of
temperature is less conclusive: it changes significantly across a
narrow temperature range $\Delta T$, from a solid low temperature
phase to a liquid high temperature phase. The behavior within $\Delta
T$ could not be resolved because of the limited time-scale
  that can be reached with simulations. There could be an abrupt
discontinuity, as in a first order transition, or there could be a
smooth change, including a temperature range with an hexatic
phase. Above $T_c$, the relaxation time increases as $T_c$ is
approached, consistent with criticality.

\acknowledgements We thank Rodrigo Arias and Patricio Cordero for useful discussions. This
work was supported by Fondecyt Grants 1100100 and 1100198, Fondap Grant 11980002, Anillo ACT 127
and a Conicyt
  2002 fellowship (M.S.). M. S. is a Howard Hughes Medical Institute
  Fellow of the Helen Hay Whitney Foundation.

\end{document}